\documentclass[onecolumn,showpacs,preprintnumbers,amsmath,amssymb]{revtex4}
\usepackage{amsfonts}
\usepackage{mathrsfs}
\usepackage{graphicx}
\usepackage{dcolumn}
\usepackage{bm}

\begin{document}

\title{Erratum: Dynamics of the Bounds of Squared Concurrence [Phys. Rev. A 79, 032306 (2009)]}

\author{Zhao Liu}%
\author{Heng Fan}
\affiliation{%
Institute of Physics, Chinese Academy of Sciences, Beijing 100190,
China
}%
\date{\today}

\maketitle In a recent paper \cite{ref1}, a shortage in our paper
was pointed out. In Eq.(15), the definition of $\eta$ should be
changed to $\eta=\min_{\{p<r\}}\omega_{p}\omega_{r}$. If in Eq.(11)
some $\omega_i=0$, Eq.(15) can only give a trivial lower bound of
$\tau(({\bf{1}}\otimes\mathcal {E})|\psi\rangle\langle\psi|)$,
namely $\tau(({\bf{1}}\otimes\mathcal
{E})|\psi\rangle\langle\psi|)\geq0$ because $\eta=0$ at this time.
However, if all $\omega_i>0$, Eq.(15) can give a non-zero lower
bound.

We also note that Eqs.(14)-(16) in our paper are derived under the
assumption that the input state $|\psi\rangle$ is expressed in the
Schmidt decomposition
$|\psi\rangle=\sum_{i}\sqrt{\omega_i}|ii\rangle$. If we consider a
general input state $|\psi\rangle=\sum_{ij}a_{ij}|ij\rangle$, we do
not know whether they hold in general, because we do not know
whether $\tau$ is local unitary invariant. However, Eq.(17) does
hold for a general input state because the tangle $\tau'$ is local
unitary invariant. Moreover, if $({\bf{1}}\otimes\mathcal
{E})|\phi^+\rangle\langle\phi^+|$ is a pure state, our Eq.(15) can
give a lower bound of concurrence itself for a general input state
$|\psi\rangle=\sum_{ij}a_{ij}|ij\rangle$, namely $\mathcal
{C}(({\bf{1}}\otimes\mathcal
{E})|\psi\rangle\langle\psi|)\geq\frac{d}{2}\sqrt{\frac{2d\eta}{d-1}}\mathcal
{C}(({\bf{1}}\otimes\mathcal
{E})|\phi^+\rangle\langle\phi^+|)\mathcal {C}(|\psi\rangle)$.

After this Erratum is published online, we find that some tighter
bounds of $\tau$, $\mathcal {C}$ and $\tau'$ can be derived from
Eq.(13) in our paper. First, we have
\begin{eqnarray}
\frac{d^{2}}{4}\eta_{min}\tau(({\bf{1}}\otimes\mathcal
{E})|\phi^{+}\rangle\langle\phi^{+}|)\mathcal
{C}^{2}(|\psi\rangle)\leq\tau(({\bf{1}}\otimes\mathcal
{E})|\psi\rangle\langle\psi|)\leq\frac{d^{2}}{4}\eta_{max}\tau(({\bf{1}}\otimes\mathcal
{E})|\phi^{+}\rangle\langle\phi^{+}|)\mathcal
{C}^{2}(|\psi\rangle),\nonumber
\end{eqnarray}
where $\eta_{max(min)}=\max(\min)_{p<r}\omega_p
\omega_r/(\sum_{i<j}\omega_i\omega_j)$ and
$|\psi\rangle=\sum_{i}\sqrt{\omega_i}|ii\rangle$. If
$({\bf{1}}\otimes\mathcal {E})|\phi^+\rangle\langle\phi^+|$ is a
pure state, we immediately have
\begin{eqnarray}
\frac{d}{2}\sqrt{\eta_{min}}\mathcal {C}(({\bf{1}}\otimes\mathcal
{E})|\phi^{+}\rangle\langle\phi^{+}|)\mathcal
{C}(|\psi\rangle)\leq\mathcal {C}(({\bf{1}}\otimes\mathcal
{E})|\psi\rangle\langle\psi|)\leq\frac{d}{2}\sqrt{\eta_{max}}\mathcal
{C}(({\bf{1}}\otimes\mathcal
{E})|\phi^{+}\rangle\langle\phi^{+}|)\mathcal
{C}(|\psi\rangle),\nonumber
\end{eqnarray}
where $|\psi\rangle=\sum_{ij}a_{ij}|ij\rangle$ is a general input
state. If $({\bf{1}}\otimes\mathcal
{E})|\phi^+\rangle\langle\phi^+|$ is a mixed state, by the technique
used in Eq.(17) in our paper, we still have
\begin{eqnarray}
\mathcal {C}(({\bf{1}}\otimes\mathcal
{E})|\psi\rangle\langle\psi|)\leq\frac{d}{2}\sqrt{\eta_{max}}\mathcal
{C}(({\bf{1}}\otimes\mathcal
{E})|\phi^{+}\rangle\langle\phi^{+}|)\mathcal
{C}(|\psi\rangle),\nonumber
\end{eqnarray}
where $|\psi\rangle=\sum_{ij}a_{ij}|ij\rangle$ is a general input
state. Of course for $\tau'$ we also have
\begin{eqnarray}
\tau'(({\bf{1}}\otimes\mathcal
{E})|\psi\rangle\langle\psi|)\leq\frac{d^2}{4}\eta_{max}\tau'(({\bf{1}}\otimes\mathcal
{E})|\phi^{+}\rangle\langle\phi^{+}|)\mathcal
{C}^2(|\psi\rangle),\nonumber
\end{eqnarray}
where $|\psi\rangle=\sum_{ij}a_{ij}|ij\rangle$ is a general input
state.

\end{document}